\documentclass[a4paper]{jpconf}

\usepackage{graphicx}
\usepackage{amsmath}
\usepackage[usenames]{color}

\usepackage{aas_macros}

 \definecolor {darkgreen}{rgb}{0.2,0.7,0.2}

\begin{document}
\title{Galactic binary science with the new LISA design}

\author{Neil Cornish and Travis Robson}

\address{eXtreme Gravity Institute, Department of Physics, Montana State University, Bozeman, MT 59717}

\ead{ncornish@montana.edu}
\ead{travis.robson@montana.edu}

\begin{abstract}
Building on the great success of the LISA Pathfinder mission, the outlines of a new LISA mission design were laid out at the $11^{\rm th}$ International LISA Symposium in Zurich. The revised design calls for three identical spacecraft forming an equilateral triangle with 2.5 million kilometer sides, and two laser links per side delivering full polarization sensitivity. With the demonstrated Pathfinder performance for the disturbance reduction system, and a well studied design for the laser metrology, it is anticipated that the new mission will have a sensitivity very close to the original LISA design. This implies that the mid-band performance, between 0.5 mHz and 3 mHz, will be limited by unresolved signals from compact binaries in our galaxy. Here we use the new LISA design  to compute updated estimates for the galactic confusion noise, the number of resolvable galactic binaries, and the accuracy to which key parameters of these systems can be measured.
\end{abstract}

\section{Introduction}

The first direct detection of gravitational waves by the advanced LIGO~\cite{PhysRevLett.116.061102} and the spectacular success of the LISA Pathfinder mission~\cite{PhysRevLett.116.231101} added a renewed energy to the $11^{\rm th}$ International LISA Symposium in Zurich. During the symposium, the LISA Consortium held a half-day meeting to begin the preparation of a formal mission proposal for the L3 launch opportunity in ESA's Cosmic Vision Program. The many decades of experience with LISA technologies and designs meant that the broad outlines of the mission quickly fell into place. The new LISA mission~\cite{LISA16} is envisioned to comprise of three identical spacecraft in a triangular formation separated by 2.5 million km in an Earth trailing orbit, with six continuously operating laser links providing heterodyne laser interferometry with ${\rm pm}/\sqrt{\rm Hz}$ sensitivity along each arm. The nominal design calls for 2 W of laser power being delivered to the optical system, with 30 cm telescopes transmitting and receiving the laser light between the spacecraft. The six laser links allow the synthesis of two Michelson-like channels that provide instantaneous measurements of the two gravitational wave polarization states, and a third Sagnac-like channel that is relatively insensitive to gravitational waves that can be used to monitor the average noise level in the detector.

\begin{figure}[htp]
	\begin{centering}
\includegraphics{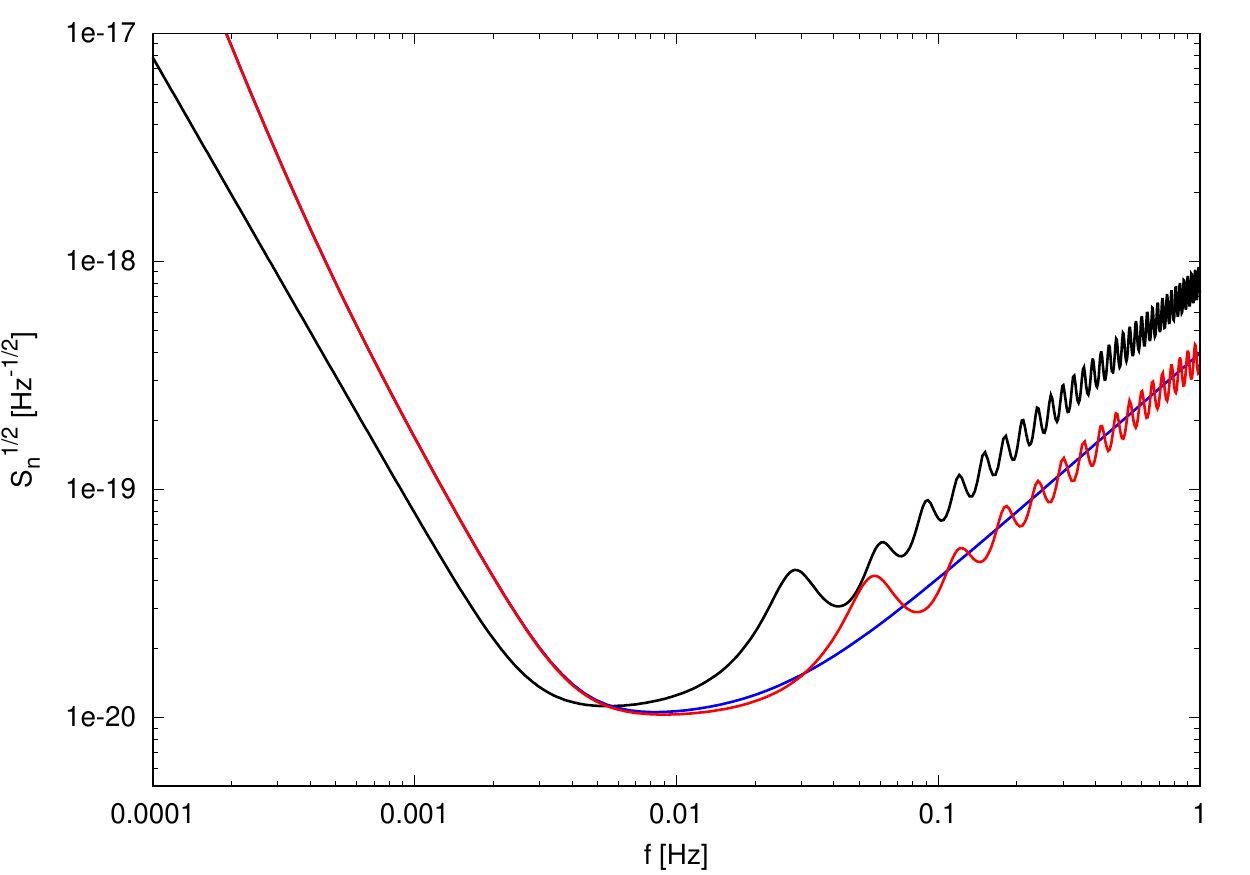} 
\caption{\label{fig:sense}  Sensitivity curves for the 1998 LISA design (black) and the new LISA design (red). Also shown is a simplified model for the sensitivity (blue), given in equation (1) of the text. The differences in the old and new sensitivity curves are almost entirely due to the reduction in arm length from 5 million km to 2.5 million km, which has the effect of reducing the low frequency sensitivity while enhancing the high frequency sensitivity.}
	\end{centering}
\end{figure}

Figure 1 compares the predicted Michelson-equivalent sensitivity of the LISA design from the 1998 pre-phase A report~\cite{prephasea98} and the new LISA design~\cite{LISA16}. Also shown is a simple approximation to the new sensitivity curve~\cite{noise16} given by
\begin{equation}
S_n(f) = \frac{20}{3}\frac{1}{L^2} \left(1 + \left( \frac{f }{1.29 f_*}\right)^2\right) \left( S_p(f) +\frac{4 S_a(f)}{(2\pi f)^4} \right).
\end{equation}
Here $f_*= c/(2 \pi L)$ is the transfer frequency, $L=2.5$ million km is the arm length, $S_p(f) = 8.9 \times 10^{-23} \; {\rm m^2 \, Hz^{-1}}$ is the white position noise, and
\begin{equation}
S_a(f) = 9\times 10^{-30} \left( 1 + 16 \left( \left(\frac{10^{-4}\, {\rm Hz}}{f}\right)^2 +  \left(\frac{2\times 10^{-5} \,{\rm Hz}}{f}\right)^{10 }\right)\right) \; {\rm m^2\, s^{-4}\, Hz^{-1}}
\end{equation}
is the colored acceleration noise. The acceleration noise model is based on a fit to the measured LISA Pathfinder performance. 

The new LISA design balances technical and budgetary constraints against a set of science requirements. A key component in this iterative design process is an assessment of the science reach of the instrument. The first step in such an assessment is to consider compact galactic binaries as these are expected to be the most numerous of the sources; so numerous in-fact that the unresolved component of the galactic population produces an effectively stochastic ``confusion noise''.  Here we provide estimates for the confusion noise level as it evolves over the lifetime of the mission, as well as estimates for the number of resolvable galactic binaries and how well key parameters, such as the orbital periods and sky location, can be determined. Our work is updates the confusion noise estimates of Nissanke {\it et al}~\cite{Nissanke:2012eh}.

\section{Galactic Confusion Noise}

The starting point for our analysis is a model for the population of compact binaries in our galaxy. We use an updated version~\cite{2012A&A...546A..70T} of the Nelemans {\it et al} \cite{2005MNRAS.356..753N} model provided to us by Valeriya Korol and Gijs Nelemans . The space density of interacting white dwarf binaries is reduced by a factor of ten relative to earlier models in response to the findings of recent observational studies\cite{2007MNRAS.382..685R,2013MNRAS.429.2143C}. 
The model predicts there are $\sim$ 26 million galactic binaries emitting in the frequency band 0.1-10 mHz. The confusion noise is estimated using the iterative subtraction procedure described in Timpano {\it et al} \cite{PhysRevD.73.122001} using a fast algorithm to generate the signals discussed in Cornish and Littenberg \cite{PhysRevD.76.083006}. The first step is to generate a realization of the instrument noise in each data channel, then co-add the signals from the population of galactic binaries. A smooth fit to the the power spectral density of the signal plus noise is found using a variant of the BayesLine algorithm~\cite{PhysRevD.91.084034}. This fit is used as an initial estimate of the effective noise spectral density, against which the signal-to-noise ratio (SNR) of each binary is computed. All binaries with ${\rm SNR} > 7$ are then subtracted from the data, and the estimate for the power spectral density of the remaining signal plus noise is computed. The SNRs of the remaining systems are re-computed using the updated (lowered) noise model, and those above threshold are subtracted. The subtraction procedure is iterated several times until few additional systems are flagged as loud. The end product is an estimate for the confusion noise and a list of resolvable sources. The mission duration has a significant impact on the level of the confusion noise. The longer the mission continues the more sources are resolved and the lower the confusion noise.

\begin{figure}[htp]
\includegraphics[clip=true,angle=0,width=0.5\textwidth]{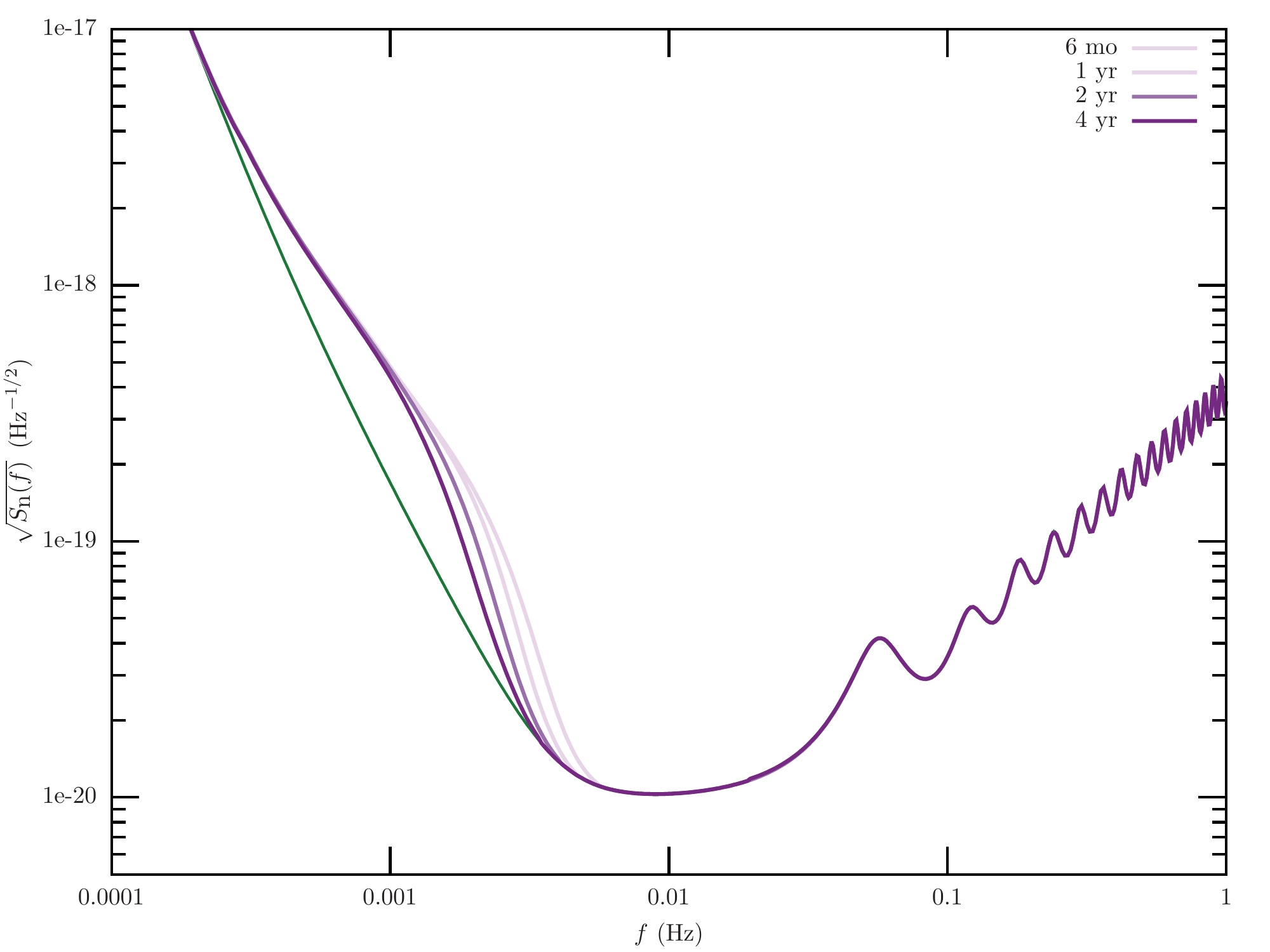}
\includegraphics[clip=true,angle=0,width=0.5\textwidth]{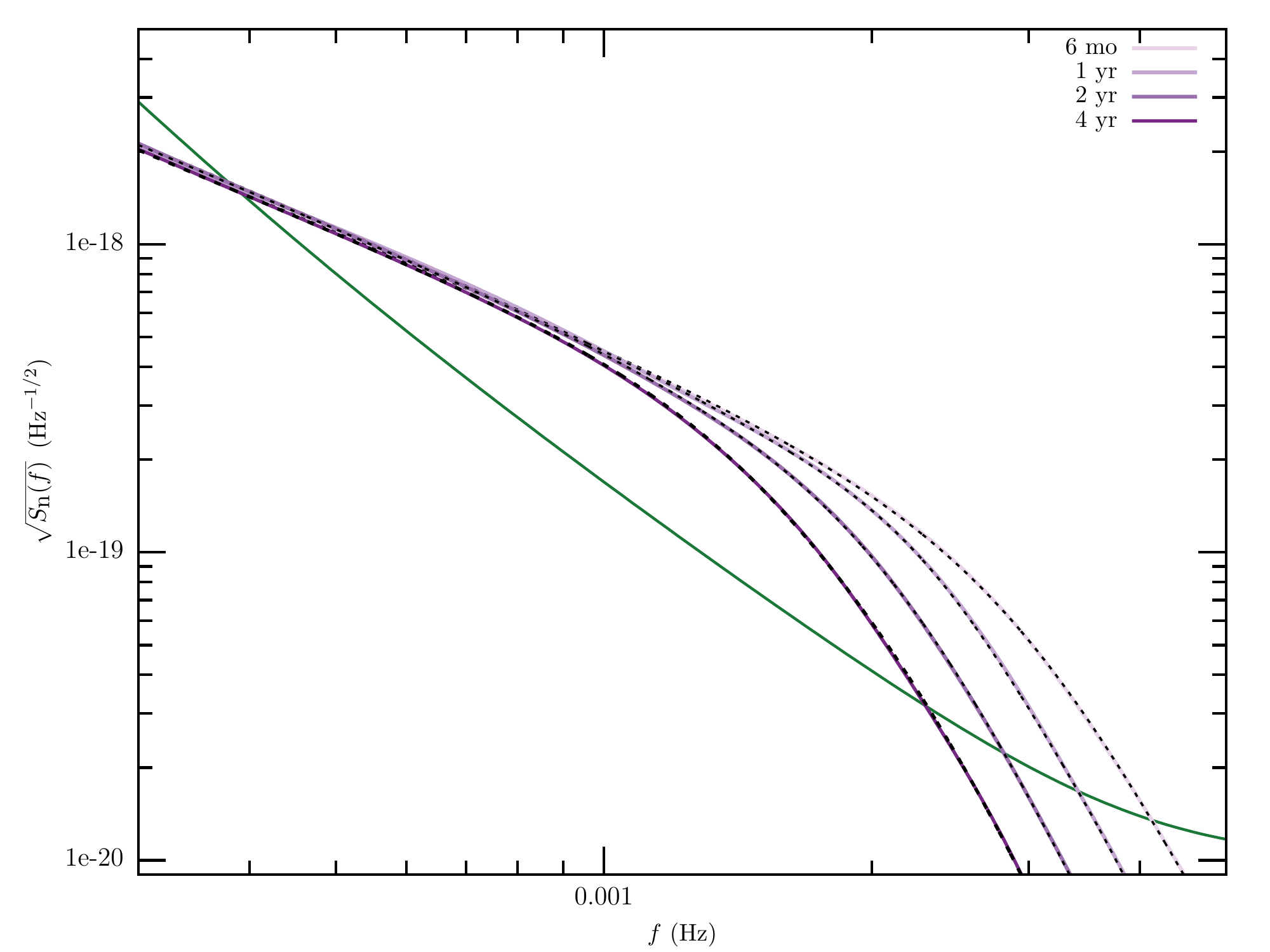}
\caption{\label{fig:confusion}  The panel on the left shows the sensitivity curve for the new LISA design including the estimated confusion noise level for a range of mission durations. The panel on the right focuses on the confusion noise impact on the sensitivity, and compares our analytic fit (black dashed lines) to the output of the simulation.}
\end{figure}

Shown in the first panel of Figure 2 are sensitivity curves that include the galactic confusion noise after 6 months, 1 year, 2 years and 4 years of data collection. The second panel in Figure 2 focuses on the confusion noise contribution, and compares our analytic fits to the results of the simulation. The analytic fits to the confusion noise have the functional form
\begin{equation}
S_c(f) = A \,  f^{- 7/3}\, e^{-f^{\alpha} + \beta  f  \sin(\kappa f) } \left[1+{\rm tanh}(\gamma(f_k-f))\right]  \, {\rm Hz}^{-1}
\end{equation}
The overall $f^{-7/3}$ slope is the theoretical prediction for the power spectrum of a population of quasi-circular binaries evolving due to gravitational wave emission. The other multiplicative factors were chosen to match the departures from the power law spectrum. The form of the fit was originally suggested by Stas Babak.  The parameters for the fits are given in Table 1. The overall amplitude and slope of there fit remain roughly constant with increasing observing time. The main change is in the knee frequency $f_k$, which gets steadily smaller with time.

\begin{table}
\begin{center}
	\begin{tabular}{ | c | c | c | c | c | }
		\hline
		& 6 mo & 1 yr & 2 yr & 4 yr \\ [0.5ex]
		\hline \hline
		%$A$ &1.80e-44 &1.86e-44 &1.95e-44& 1.84e-44 \\
		$\alpha$&0.133&0.171&0.165&0.138\\
		$\beta$&243&292&299&-221 \\
		$\kappa$&482&1020&611&521\\
		$\gamma$&917&1680&1340&1680\\
		$f_{k}$&0.00258&0.00215&0.00173&0.00113\\
		\hline\hline
	\end{tabular}
\end{center}
\caption{Parameters of the analytic fit the Galactic confusion noise as described by equation (1). The amplitude $A$ has been fixed to $1.80e-44$. Note that the knee frequency $f_{k}$ decreases with observation time as expected from Figure 2. Similarly, $\gamma$ increase with observation time to leading to steeper drop off in confusion noise  with increasing frequency.}
\end{table}

\section{Galactic Binary Science}

While the unresolved galactic binaries are a source of noise that negatively impacts the science that can be done with other sources, the resolvable galactic binaries tell a rich story about stellar evolution and the distribution of stars in the galaxy.  The distribution of resolved signals as a function of orbital period can be used to constrain population synthesis models. The orbital period is very well measured even for systems at the detection threshold. Figure 3 shows the number density of resolved signals per $1/T_{\rm obs}$ frequency bin for a 4 year mission using both Michelson-like data channels.

\begin{figure}[htp]
	\begin{centering}
		\includegraphics[clip=true,angle=0,width=0.7\textwidth]{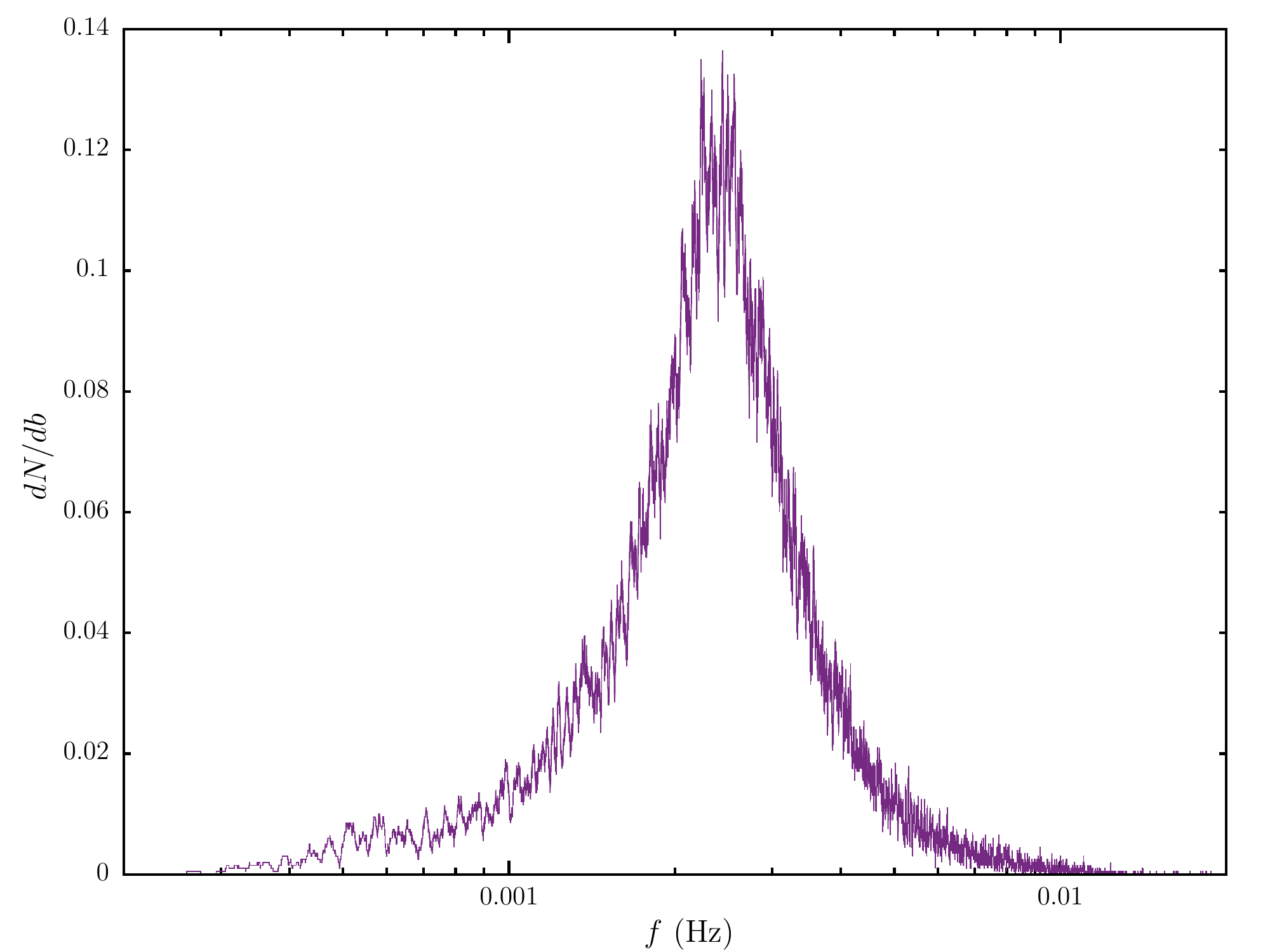} 
		\caption{\label{fig:denisty}  The number density of resolved signals per frequency bin for a 4 year mission using both Michelson-like data channels. Note that the density of resolved signals is largest between 2 - 3 mHz, which is where the Galactic confusion noise changes the most over time as shown in Figure 2.}
	\end{centering}
\end{figure}

A sub-set of the resolved signals will be well localized on the sky, and a smaller sub-set will undergo sufficient frequency evolution to allow measurement of the chirp mass and distance. Table 2 lists the predicted number of detected systems, along with the number that are well localized in sky position and distance, and the number for which we can measure the chirp mass. The parameter measurement accuracy is estimated using the Fisher information matrix. The limitations of this approach are well known, but we expect the estimates given in Table 2 would be close to those found using a full Bayesian analysis since the results are for the best measured systems, which typically have high SNR. 

\begin{table}
	\begin{center}
		\begin{tabular}{ | c | c | c | c | c | }
			\hline
			&6 mo & 1 yr & 2 yr & 4 yr \\ [0.5ex]
			\hline\hline
			\# detected &6,590&11,142&18,281&29,059\\
			2D mapped &104&1,065&4,138&6,304\\
			3D mapped &19&129&1,010&2,373\\
			$\mathcal{M}$ measured &233&737&4,432&10,770\\
			\hline\hline
		\end{tabular}
	\end{center}
	\caption{The first row indicates the number of sources detected using the two Michelson-like data channels for 6 months, 1 year, 2 year, and 4 year observation periods. If a source is localized to better than a square degree we deemed it to be mapped on the sky (2D mapped). Additionally, if its distance has also been obtained to better than 10\% we deemed to have been mapped in 3D. The last row indicates the number of sources whose chirp mass $\mathcal{M}$ have been determined to better than 20\%.}
\end{table}

\begin{figure}[htp]
	\begin{centering}
		\includegraphics[clip=true,angle=0,width=0.7\textwidth]{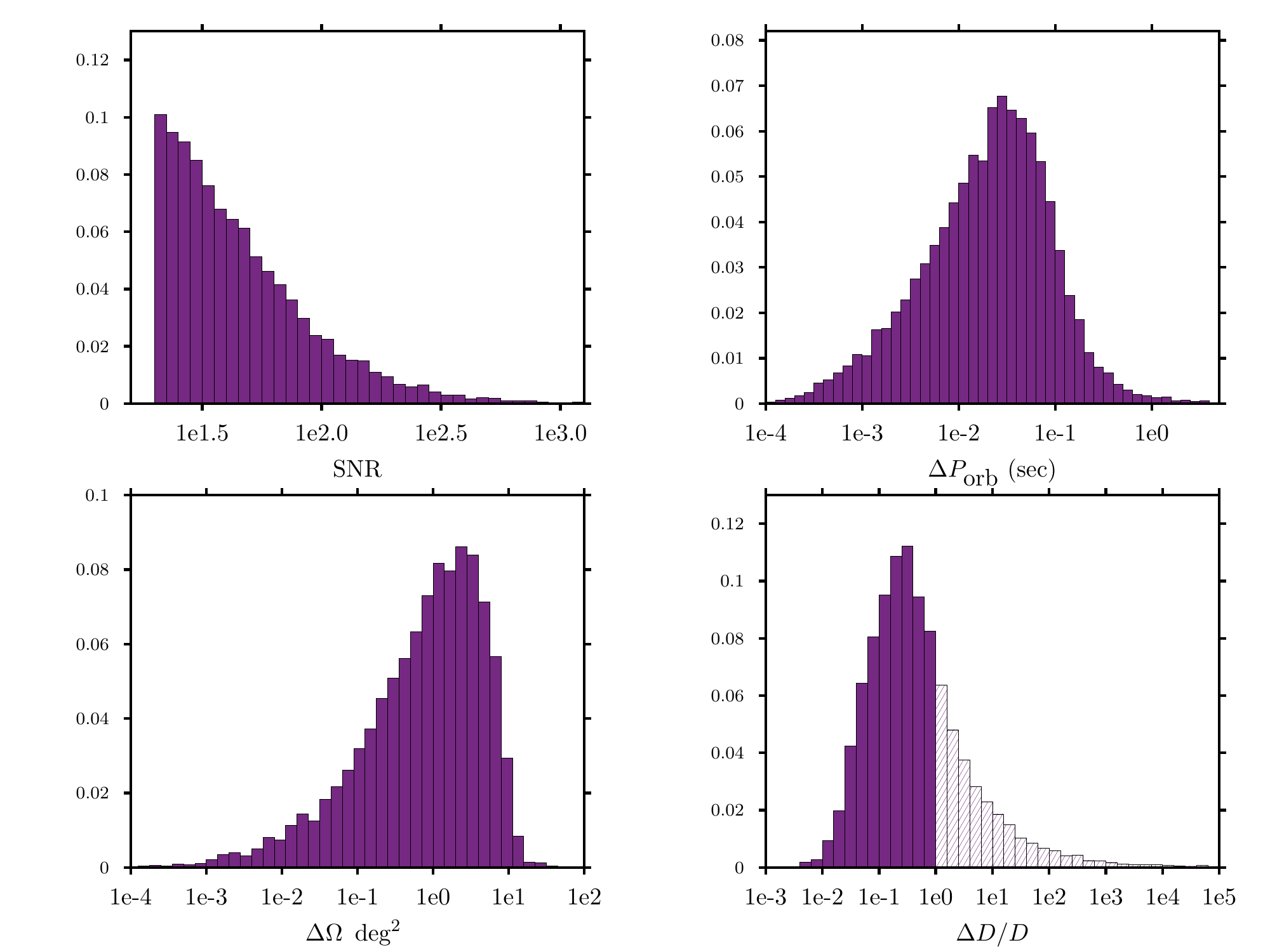} 
		\caption{\label{fig:loud_histograms}  Histograms for the SNR and parameter estimation errors for sources with SNR $> 20$ using the two Michelson-like data channels for 4 year observation period. The orbital period (upper right) is determined to better than a second accuracy, reflecting the precision to which we expect to measure the galactic binary frequencies. In the lower left panel we see that the angular resolution of many sources is determined to within a square degree. Distance is the hardest parameter to extract, as depicted by the lower right panel. Results which are nonphysical (i.e. $\Delta D/D > 1$) are dashed. }
	\end{centering}
\end{figure}

Figure 4 shows histograms of the SNR, and the expected accuracy in the measurement of orbital period, sky location and distance for systems with ${\rm SNR} > 20$,  assuming a 4 year mission with 6 links. The limitations of the Fisher matrix approximation for estimating parameter estimation motivated the higher SNR cut.

\begin{figure}[htp]
	\begin{centering}
		\includegraphics[clip=true,angle=0,width=0.7\textwidth]{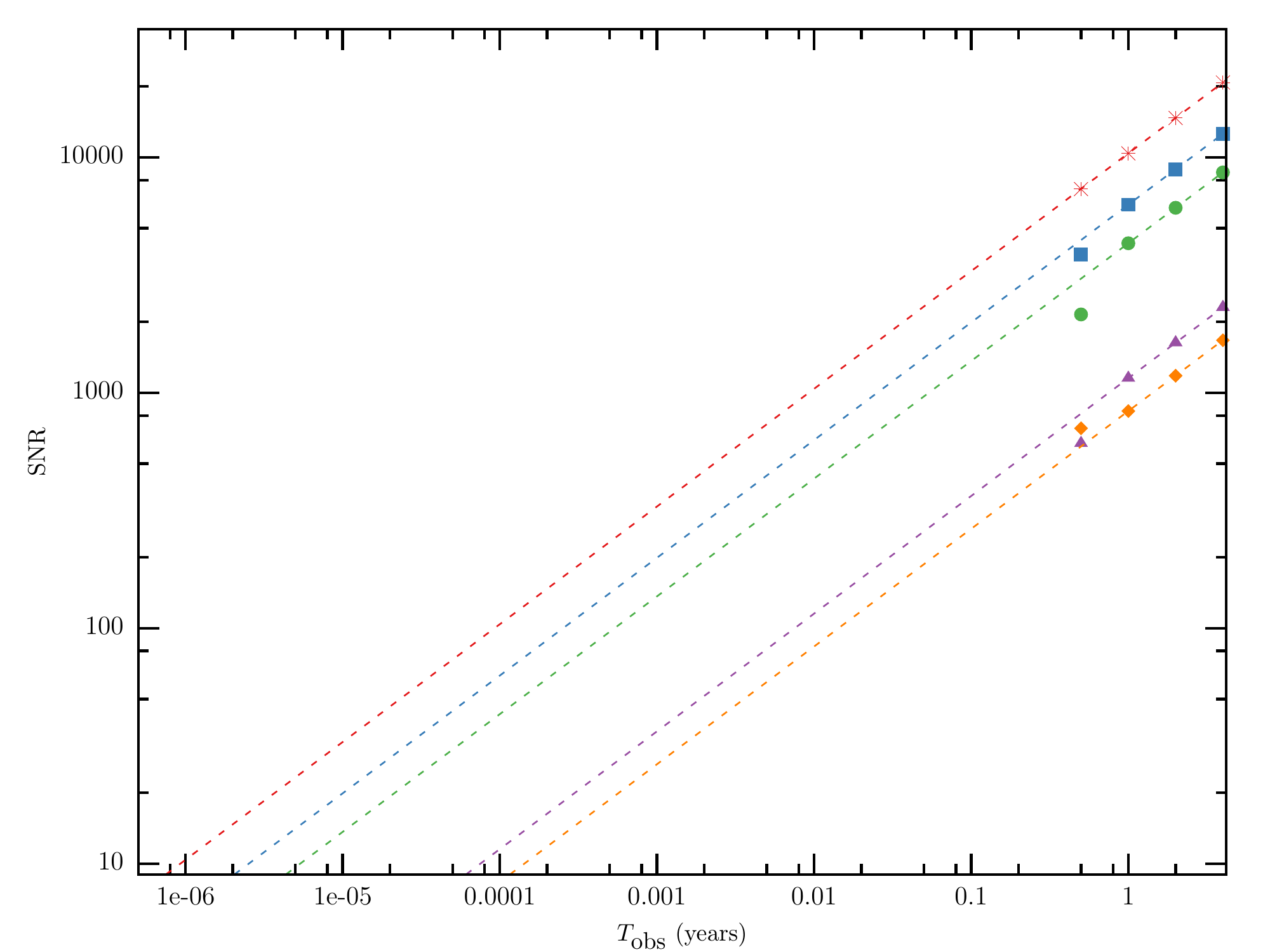} 
		\caption{\label{fig:SNR}  The signal-to-noise ratios for the five loudest galactic binaries as a function of observation time. The dashed lines show the $T_{\rm obs}^{1/2}$ growth in SNR that we expect for a fixed noise level. At early times the confusion noise level is dropping rapidly form many of the brightest sources so the SNR growth is faster.}
	\end{centering}
\end{figure}

The SNR distribution has a long tail that extends to very high SNR. These very high SNR systems are of particular interest as they will typically be very well localized (especially the higher frequency systems), making them good candidates for joint electromagnetic-gravitational observation~\cite{Littenberg:2012vs}, and they will also be among the first sources detected by LISA. Figure 6 plots the SNR of the five brightest galactic binaries as a function of mission duration. The dashed lines indicate the $T_{\rm obs}^{1/2}$ growth in SNR that we expect if the noise level is constant. For most galactic binaries the SNR grows faster than this since the effective noise level drops as weaker signals are resolved and removed. Extrapolating the curves back below 6 months is not  entirely justified since the confusion noise is non-stationary and varies on a 6 month period, and the SNR does not precisely follow the $T_{\rm obs}^{1/2}$ scaling, but even with the caveats, we see that the loudest signals will be detectable within hours of starting data collection.

\section{Discussion}
Using the current best estimates for the galactic binary population we computed the galactic binary confusion noise for the new LISA design and found that is the dominating source of noise in the frequency band $0.5-3.0$ mHz. Fits for the confusion noise curves are been provided. Additionally we provided estimates for how many galactic binary sources will be detected, and how well we expect to be able to estimate their parameters. Determining the 3D position will be the most challenging; many sources will be located to better than a square degree on the sky, but breaking the degeneracy between chirp mass and distance hurts distance estimation. We found that most sources will be resolved above $3$ mHz, with orbital periods extracted to better than one second accuracy. 

\ack
We wish to thank the organizers of the $11^{\rm th}$ International LISA Symposium for putting together a memorable meeting, and our colleagues in the LISA Consortium for their work on the mission design. We are grateful for the support provided by NASA grant NNX16AB98G.

\section{References}

\bibliography{reference.bib}

\end{document}